\documentclass[%
preprint,
superscriptaddress,
showpacs,
amsmath,amssymb,
aps,
prb,
floatfix,
]{revtex4-1}

\usepackage{graphicx}
\usepackage{dcolumn}
\usepackage{bm}
\usepackage{multirow}
\usepackage{xspace}

\begin{document}

\newcommand{\La}{$^{139}$La\xspace}
\newcommand{\Al}{$^{27}$Al\xspace}
\newcommand{\C}{LaCuAl$_3$\xspace}
\newcommand{\A}{LaAuAl$_3$\xspace}
\newcommand{\V}{$V_{zz}$\xspace}

\title{Local atomic arrangement in LaCuAl$_3$ and LaAuAl$_3$ by NMR and density functional theory}

\author{Vojt\v{e}ch Chlan}
\affiliation{Charles University, Faculty of Mathematics and Physics, Department of Low Temperature Physics, V Hole\v{s}ovi\v{c}k\'ach 2, 180 00 Prague 8, Czech Republic}
\author{Petr Dole\v{z}al}
\affiliation{Charles University, Faculty of Mathematics and Physics, Department of Condensed Matter Physics, Ke Karlovu 5, 121 16 Prague 2, Czech Republic.}
\author{R\'{a}chel Sgallov\'{a}}
\affiliation{Charles University, Faculty of Mathematics and Physics, Department of Low Temperature Physics, V Hole\v{s}ovi\v{c}k\'ach 2, 180 00 Prague 8, Czech Republic}
\author{Christian Franz}
\affiliation{Physik Department, Technische Universität München, D-85747, Garching, Germany}
\affiliation{Heinz Maier-Leibnitz Zentrum (MLZ), Technische Universität München, D-85748 Garching, Germany}
\author{Pavel Javorsk\'{y}}
\affiliation{Charles University, Faculty of Mathematics and Physics, Department of Condensed Matter Physics, Ke Karlovu 5, 121 16 Prague 2, Czech Republic.}

\begin{abstract}
CeCuAl$_3$ and CeAuAl$_3$, crystallizing in the non-centrosymmetric BaNiSn$_3$ tetragonal structure, are known mainly for their unusual neutron scattering spectra involving additional excitations ascribed to vibron quasi-bound quantum state in CeCuAl$_3$ and anti-crossing of phonon and crystal field excitations in CeAuAl$_3$. In this work, we present results of nuclear magnetic resonance experiments on their lanthanum analogues -- \C and \A. The character of nuclear magnetic resonance spectra of \La, \Al, and $^{65}$Cu measured in \A and \C is dominated by electric quadrupole interaction. The spectral parameters acquired from experimental data are confronted with values obtained from the electronic structure calculations. The results show remarkable differences for the two compounds. The \La spectrum in \A can be interpreted by a single spectral component corresponding to uniform environment of La atoms in the crystal structure, whereas for \C the spectrum decomposition yields a wide distribution of spectral parameters, which is not possible to explain by a single La environment, and multiple non-equivalent La positions in the crystal structure are required to interpret the spectrum.
\end{abstract}

\maketitle
\section{Introduction}

LaCuAl$_3$ and LaAuAl$_3$ belong to a large group of $RTX_3$ ($R$ stands for rare earth atoms, $T$ for transition metal, and $X$ for a p-metal) compounds crystallizing in the tetragonal BaNiSn$_3$-type structure. The structure is non-centrosymmetric and the lack of inversion symmetry affects physical properties of these materials.
The most intriguing properties exhibit the Ce-based ones. Remarkable is the observation\cite{CeTSi3} of the pressure-induced superconductivity in Ce$T$Si$_3$ (T = Co, Rh, Ir)  which classifies these materials among superconductors where the properties are dictated by an antisymmetric spin-orbit coupling as a consequence of the lack of inversion symmetry. In compounds with $X$ = Al, the superconductivity was not found until now.
Concerning compounds of our present study, \C and \A remain paramagnetic down to 0.4 K and their Ce-based counterparts CeCuAl$_3$ and CeAuAl$_3$ order antiferromagnetically at low temperatures ($T_N = 2.5$~K for CeCuAl$_3$, see Ref.~\onlinecite{CeCuAl3magstr}, and $T_N = 1.1$~K for CeAuAl$_3$, see Ref.~\onlinecite{CeAuAl3magstr}).

The main interest in CeCuAl$_3$ and CeAuAl$_3$ lies in experimental observations which were ascribed to electron-phonon interaction. The interaction between crystal electric field (CEF) and lattice degrees of freedom, usually considered as a weak effect in intermetallic compounds, is believed to be responsible for unusual phenomena experimentally observed in these two compounds by inelastic neutron scattering. The anti-crossing of the CEF level and the phonon states was recently reported in CeAuAl$_3$.\cite{Cermak} CeCuAl$_3$ then belongs to the few compounds which show an additional peak in the neutron energy spectra \cite{AdrojaPRL} which cannot be explained on the basis of standard CEF theory. This effect, first observed in the cubic CeAl$_2$,\cite{CeAl2Steglich, CeAl2Loew} was theoretically described on the basis of a strong magnetoelastic interaction between the CEF excitations and lattice vibrations which leads to a formation of a new quantum state -- vibron quasi-bound state.\cite{Thalmeier}

The alternative view on the existence of the additional excitation in the neutron spectra of CuCuAl$_3$ was discussed on the basis of certain Cu-Al atomic site disorder which would lead to existence of different local surroundings of the Ce ions.\cite{Vlaskova} Although the X-ray single crystal study\cite{CeCuAl3struct} concluded the BaNiSn$_3$-type of structure to be the most reliable one for CeCuAl$_3$, the crystal structure is apparently not perfectly ordered one, but there is indeed a certain, relatively low Cu-Al atomic site disorder as reported in several studies.\cite{Kawamura,Franz} Such disorder occurs also in other members of this family of materials and has a huge impact on their electronic properties. For example, the type of magnetic order in CeCuGa$_3$ was discussed to be dependent on the Cu-Ga atomic site
disorder: the ordered non-centrosymmetric BaNiSn$_3$-type variant tends to be antiferromagnetic, while a ferromagnetic order is reported for the disordered BaAl$_4$-type structure. On the other hand, no such disorder was found for CeAuAl$_3$ or LaAuAl$_3$.\cite{Franz}

As the atomic site disorder might substantially influence the electronic properties of Ce$T$Al$_3$ compounds, it is important to study it more deeply by microscopic techniques. The nuclear magnetic resonance (NMR) spectroscopy is the most powerful technique sensitive to the local atomic surroundings. The direct investigation of the Ce-based compounds by NMR is, however, complicated by lack of cerium isotopes with nuclear spin, and additionally, the resonance lines would presumably be significantly broadened by interaction with the atomic magnetic moments. Therefore, we have performed a comparative NMR study of the \C and \A compounds. The structural properties of the Ce$T$Al$_3$ and La$T$Al$_3$ are very similar (see, e.g., Ref.\ \onlinecite{CeCuAuAl3}), so we assume that the main observations found for the lanthanum compounds are valid also for their cerium analogs.

\section{Methods}

Polycrystalline samples of \C and \A were prepared by arc-melting stoichiometric mixtures of pure elements (Solid State Electrotransport purified La, 6N Cu, 5N Au and 6N Al) in mono-arc furnace under protection of argon atmosphere. The samples were turned and re-melted several times to achieve better homogeneity. Additionally, the samples were sealed under vacuum in a quartz glass and annealed for 10 days at 850~$^{\circ}\textrm{C}$ -- they were heated up to 850~$^{\circ}\textrm{C}$ with the rate of 3~$^{\circ}\textrm{C}$ per minute and cooled down with the rate of 1~$^{\circ}\textrm{C}$ per minute. The crystal structure and phase purity was checked by powder X-ray diffraction and scanning electron microscope. The powder samples for the NMR experiments were sieved by 50~$\mu$m sieve.

Nuclear magnetic resonance spectra of \La, \Al, and $^{65}$Cu nuclei were measured by Bruker Avance II spectrometer at room temperature in 9.4~T static magnetic field. The frequency-swept spectra were acquired by a modified Carr-Purcell-Meiboom-Gill (CPMG) pulse sequence where all (up to 50) induced spin echoes  were coherently summed and their sum Fourier transformed. The frequency steps were varied (10--250 kHz) in order to cover all spectral features properly, the NMR probe was well tuned and matched at each excitation frequency, and the rf pulse lengths were set to induce maximum intensity of the signal. Typical rf pulse lengths were 3--6~$\mu$s, the delays between pulses in the CPMG train were 70--150~$\mu$s, and 1000--2000 scans were accumulated at each excitation frequency with delays between the scans 50--300~ms depending on spin-lattice relaxation.

In the presence of a strong electric quadrupole interaction its influence cannot be neglected within duration of the excitation rf pulse and a mixing of populations during the pulse occurs -- in contrast to situation with only a magnetic Zeeman interaction of nuclear spin $I$ with static magnetic field, when each energy level corresponds to a well-defined quantum number $m \in \left<-I,I\right>$. Different transitions between the quadrupolarly split nuclear energy levels then possess different nutation frequencies, and therefore, also different lengths of the optimal rf pulse.\cite{Freude} The excitation by a CPMG sequence with fixed rf pulse lengths thus fails to accomplish optimal excitation, and as a consequence, the relative contributions of individual transition to the overall intensity may differ from the ratios expected from theory. We also note that our approach in addition neglects phase in the evaluation of spectral intensity, i.e., plotting magnitude $\sqrt{(\chi')^2 + (\chi'')^2 }$ vs.\ frequency, which results in a small perturbation of the powder pattern shape compared to the case where the absorption component of the susceptibility $\chi'$ is used. This approach, however, had to be chosen due to extreme breadths of the NMR spectra and in order to obtain reasonable signal-to-noise ratio.

The electronic structures of \A and \C with various atomic arrangements were calculated within the density functional theory (DFT) by means of full-potential all-electrons augmented plane wave + local orbitals method implemented in WIEN2k code.\cite{wien2k} The radii of atomic spheres were 2.5~a.u.\ for La and Au, 2.1~a.u.\ for Cu, and 2.0~a.u.\ for Al. The lattice parameters as well as the atomic coordinates were optimized (within the space group symmetry of each calculated structure) by minimizing the total energy and atomic forces. The size of the basis set and the density of the \textit{k}-point mesh were well converged with respect to the structure geometry (lattice parameters) and the calculated quantities of interest (electric field gradients). The value for the RK$_\mathrm{max}$ parameter was 8.0, yielding the matrix size about 1600 basis functions (half the size for the structures with \textit{I}-centration). 3000 \textit{k}-points were used, leading to 220--400 \textit{k}-points in the irreducible part of the Brillouin zone. The charge density and potentials were Fourier expanded up to the largest $k$-vector $G_\mathrm{max}$ = 14~$\sqrt{\mathrm{Ry}}$ and as the exchange-correlation potential Perdew-Burke-Ernzerhof variant\cite{ggapbe} of generalized gradient approximation was used.

\section{Results and Discussion}

\begin{figure}
	\includegraphics[width=\columnwidth]{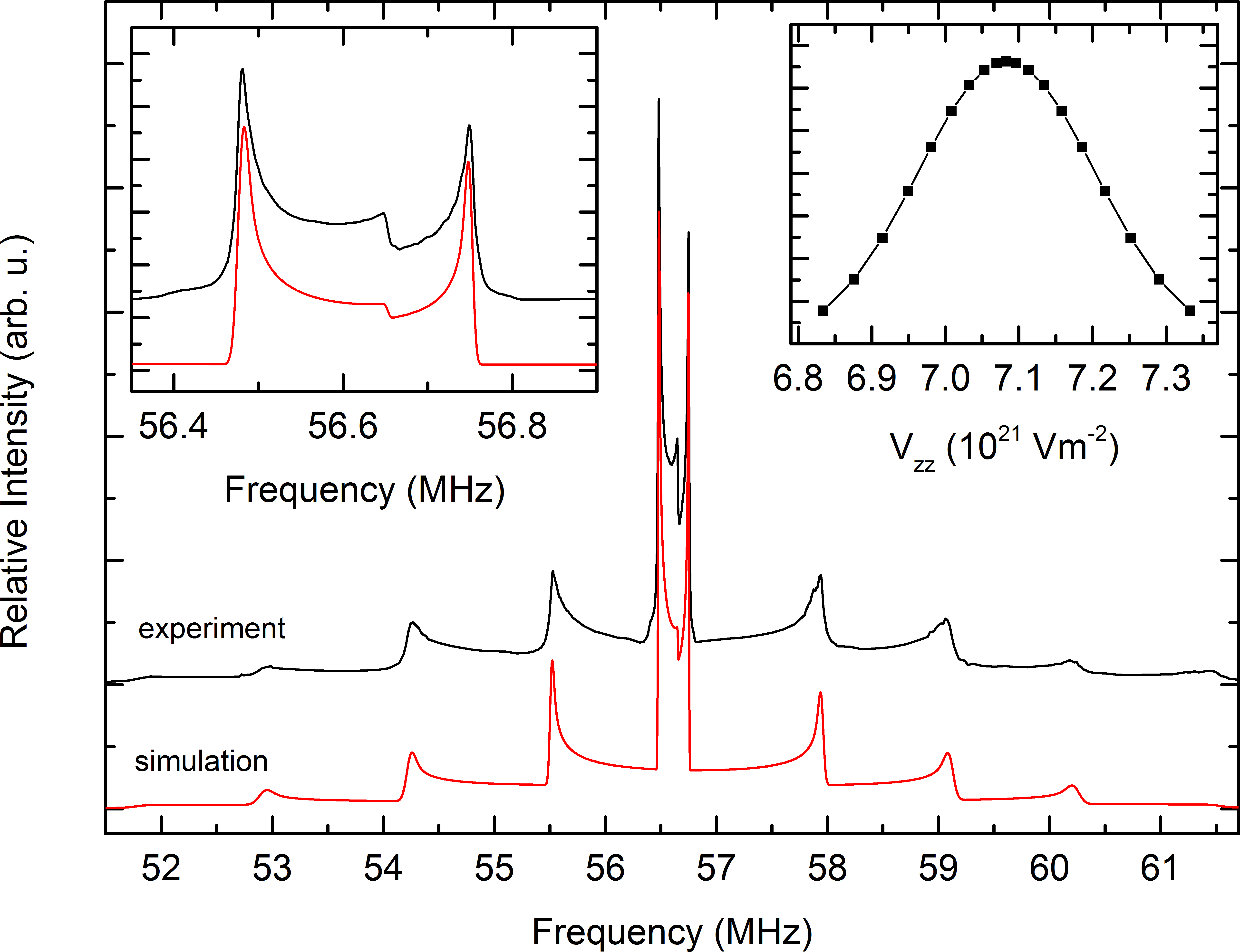}
	\caption{\label{fig:LaA} \La NMR spectrum of \A in comparison with simulated spectrum. The inset on the left side details the central transition, inset on the right side shows distribution of \V parameter used in the simulation.}
\end{figure}

\begin{figure}
	\includegraphics[width=\columnwidth]{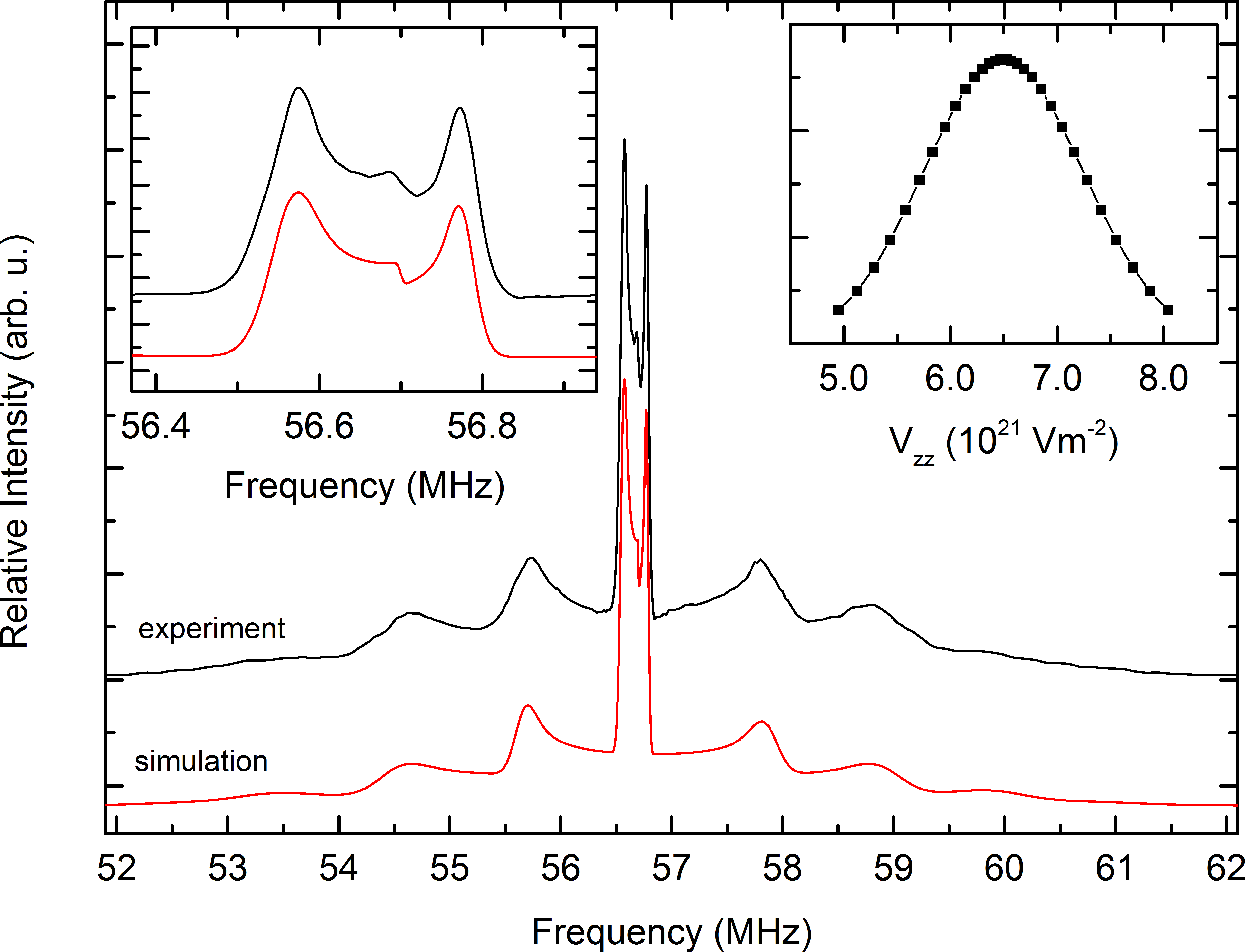}
	\caption{\label{fig:LaC} \La NMR spectrum of \C in comparison with simulated spectrum. The inset on the left side details the central transition, inset on the right side shows distribution of \V parameter used in the simulation.}
\end{figure}

Measured \La NMR spectrum of \A (see Fig.~\ref{fig:LaA}) shows features which are characteristic for a powder spectrum in presence of a strong electric quadrupole interaction of the nuclear quadrupole momentum $Q$ with electric field gradient (EFG). For \La nucleus (spin $I$=7/2) there are seven nuclear transitions expected: one central transition located approximately at the Larmor frequency $\nu_L\sim~56.53$~MHz and six satellite transitions spread in frequency range $\sim$50--64~MHz. Due to a random orientation of the powder grains, each nuclear transition contributes with all possible mutual orientations of the static magnetic field vector and the EFG tensor. The quadrupole interaction is often considered as a perturbation to the Zeeman interaction with static magnetic field, in the first-order expansion ("first-order quadrupole correction") the frequencies of the transitions $m$ $\leftrightarrow$ $m+1$ can be expressed as:\cite{PPMan}
\begin{widetext}
\begin{eqnarray}
\omega_{m,m+1} &=& \frac{\omega_Q}{2} (1-2m)(3\cos^2 \vartheta-1+\eta\sin^2 \vartheta \cos 2\phi), \\
\omega_Q &=& \frac{3eQV_{zz}}{2I(2I-1)\hbar}, \nonumber
\end{eqnarray}
\end{widetext}
where $\omega_Q$ is the quadrupole frequency, \V is the $zz$ component of the EFG tensor, $\eta$ is the parameter of asymmetry $\eta = \frac{V_{xx}-V_{yy}}{V_{zz}}$, and $\vartheta$ and $\varphi$ are the Euler angles of the external magnetic field within the principal axis system of the EFG tensor. Since $\omega_{-\frac{1}{2},\frac{1}{2}} = 0$, the central transition $-\frac{1}{2}$ $\leftrightarrow$ $\frac{1}{2}$  is not affected by the first-order quadrupole correction. In order to describe the effect of strong quadrupole interaction on the central transition, one has to expand up to the second-order ("second-order quadrupole correction"):\cite{PPMan}
\begin{widetext}
\begin{eqnarray}
\omega_{-\frac{1}{2},\frac{1}{2}} &=& -\frac{\omega_Q^2}{6\omega_L} \left(I(I+1)-\frac{3}{4}\right)\left(A(\varphi,\eta)\cos^4 \vartheta +B(\varphi,\eta)\cos^2 \vartheta +C(\varphi,\eta)\right), 
\end{eqnarray}
\end{widetext}
where A, B, and C terms possess the dependence on angle $\varphi$ and asymmetry $\eta$ (their expressions can be found in Ref.~\onlinecite{PPMan}).

The experimental spectrum of \A in Fig.~\ref{fig:LaA} clearly has a single dominant 7/2-spin powder pattern, which corresponds to one prevailing value of \V. This is evident from comparison with  \La spectrum simulated using a narrow distribution of \V around value $7.1 \times 10^{21}$ Vm$^{-2}$, with $\eta=0$, and convoluted with a Gaussian peak shape of width 6~kHz. Such simulation sufficiently describes all the spectral features of the experiment. The need for a distribution of \V rather than a single value is dictated by the fact that the quadrupole splitting for the satellite transitions is much more sensitive to \V (first-order effect), unlike for the central transition (second-order effect). A mere convolution with Gaussian shape would not capture the breadths of the central and satellite transitions all at once.

In the experimental \La  NMR spectrum of \C (see Fig.~\ref{fig:LaC}) the powder pattern spectral features are smeared-out in contrast to \A, as is documented by an order of magnitude broader distribution of \V in the respective simulation (note the different scales of the \V-axes in the insets of Figures~\ref{fig:LaA} and \ref{fig:LaC}). Qualitatively, based solely on the experimental data, we can thus interpret the \La spectra in the following way: lanthanum in the \A experiences very homogeneous environment with axial symmetry (asymmetry $\eta=0$). On the contrary, in \C the broad range of \V indicates a mixture of different lanthanum sites. 

\begin{table}
	\caption{\label{tab:structs}Symmetries of modeled crystal structures of \A and \C, and calculated EFG parameters of \La.}
	\begin{ruledtabular}
		\begin{tabular}{lccdddd}
			\multirow{2}{*}{Structure} & space  & La-site& \multicolumn{2}{c}{LaAuAl$_3$} &\multicolumn{2}{c}{LaCuAl$_3$}  \\
			&  group &  point group &  \multicolumn{1}{c}{$V_{zz}$} & \multicolumn{1}{c}{$\eta$} & \multicolumn{1}{c}{$V_{zz}$} & \multicolumn{1}{c}{$\eta$}   \\
			\colrule\rule{0pt}{9pt}\textit{str}01 &	\textit{I}4\textit{mm} &	4\textit{mm} &		6.91	& 0& 	6.36	&0 \\
			\textit{str}02 &		\textit{P}4/\textit{mmm} &	4/\textit{mmm} & 5.65 & 0 & 4.80 & 0 \\
			& &  4/\textit{mmm} & 7.98 & 0 & 8.28 & 0 \\
			\textit{str}03 &	\textit{P}4/\textit{nmm}	& 4\textit{mm} &		7.47	&0&	6.95&	0 \\
			\textit{str}04 &	\textit{P}/\textit{mm}2 &	\textit{mm}2 & -3.29 & 0.97 & 4.25 & 0.93 \\
			& &  \textit{mm}2 & 7.57 & 0.21 & 6.02 & 0.02 \\
			\textit{str}05 &			\textit{P}/\textit{mm}2 &	\textit{mm}2 & 4.25 & 0.44 & 4.32 & 0.72 \\
			& &  \textit{mm}2 & 6.17 & 0.32 & 6.49 & 0.48 \\
			\textit{str}06 &	\textit{P}42/\textit{mmc}&	\textit{mmm}&		-3.06&	0.76&	-4.90&	0.59 \\
			\textit{str}07 &	\textit{I}$\bar{4}$\textit{m}2&	$\bar{4}$2\textit{m} &		2.68 &	0	&3.80	&0 \\
			\textit{str}08 &	\textit{P}4/\textit{nmm}&	4\textit{mm}&		1.25&	0	&2.65&	0 \\
		\end{tabular}
	\end{ruledtabular}
\end{table}

In order to assess our interpretation it is required to show that (i) different values of \V can be attributed to different arrangements of Au/Al or Cu/Al atoms in the structure and that (ii) there are no other effects that could possibly explain the smeared-out shape of \La NMR spectrum in \C. Therefore, we compare the experiment with results of DFT calculations where the ordering of Au and Al atoms (or Cu and Al, respectively) was modeled within one tetragonal unit cell containing ten atoms, i.e., La$_2$Au$_2$Al$_6$ and La$_2$Cu$_2$Al$_6$. The starting point was the \textit{I}4\textit{mm} BaAl$_4$-type structure with La in the 2\textit{a} sites and with Au (Cu) and Al atoms to be arbitrarily allocated within all the remaining 2\textit{a} and 4\textit{b} sites. Such model allows for eight crystallographically nonequivalent structures (listed in Table~\ref{tab:structs}), which differ by symmetry, atomic arrangement, and most importantly also by the calculated EFG parameters at La sites.

\begin{figure}
	\includegraphics[width=\columnwidth]{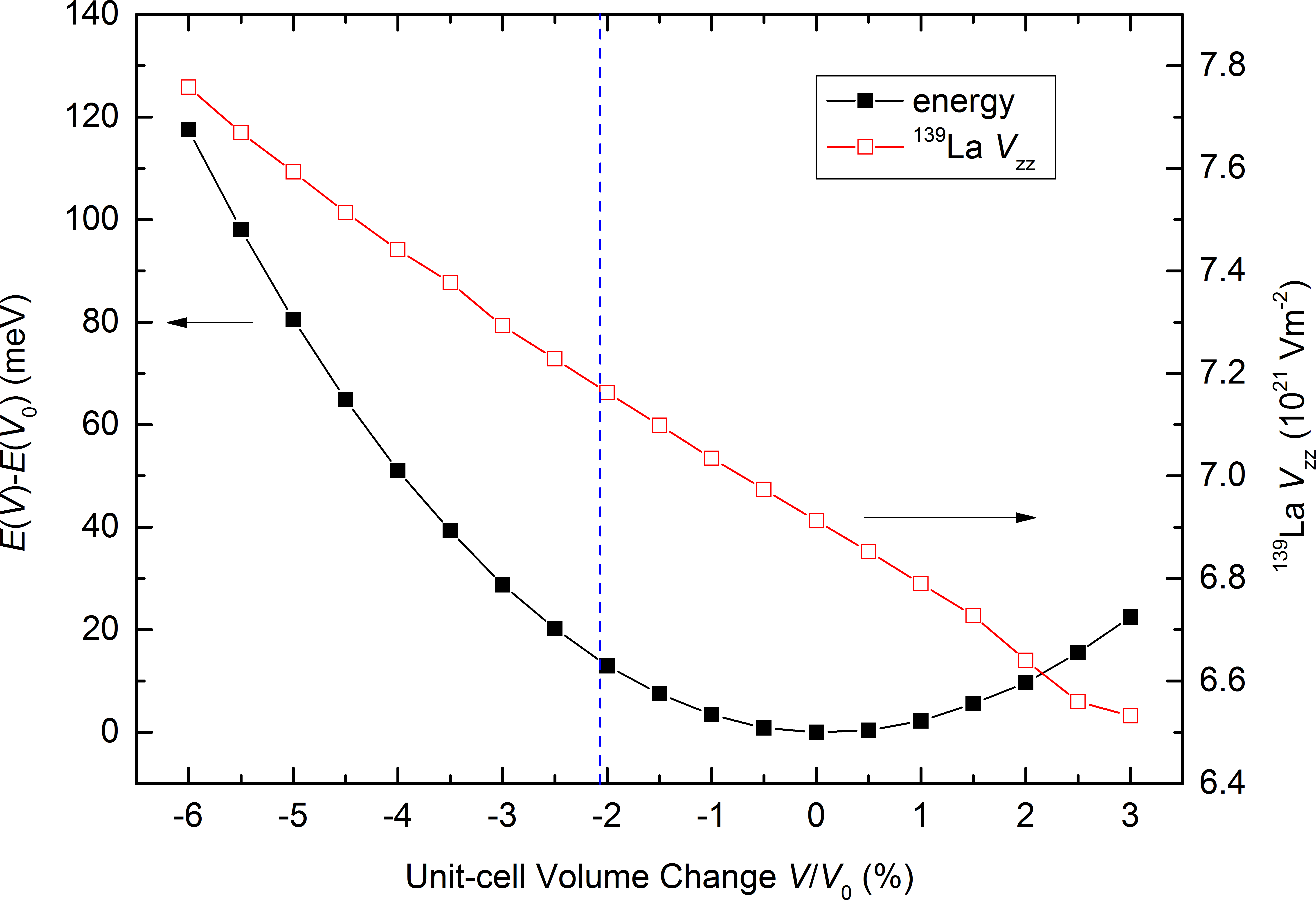}
	\caption{\label{fig:vol} Volume dependence of the total energy and \V on La nucleus in \A \textit{str}01 structure. Blue dashed line denotes unit cell volume corresponding to the room temperature experiment.\cite{LaCuAl3lattpars}}
\end{figure}

The calculated values of \V in Table~\ref{tab:structs} vary substantially for different atomic configurations; likewise the value of $\eta$ corresponds to presence of local symmetry axis. For the calculated values to be directly comparable with room temperature experiments, one should increase the calculated values by a few percent, in order to reflect the difference between unit cell volume in calculation and in experiment. This volume discrepancy is expected and caused by two sources: first, the calculation relates to temperature 0~K and as such corresponds to approximately 1--2~\% smaller unit cell volume due to thermal expansion. Second, the exchange-correlation potential\cite{ggapbe} used in the calculations overestimates\cite{volpot1,volpot2,volpot3} the equilibrium volume by additional approximately 0--2~\%. The value of \V on La nucleus is dominated by p-p and to some extent also d-d contributions which increase in value with decreasing volume as the La 5p and 4d orbitals become compressed towards the nucleus (see the volume dependence in Fig.~\ref{fig:vol}). The calculated equilibrium volume is about 2~\% larger and one should thus increase the calculated \V by roughly 5 \% for comparison with experiment.

Confronting the calculated \V values for \A in Table~\ref{tab:structs} with the \V distribution acquired from the NMR experiment (Fig.~\ref{fig:LaA}), clearly only the \textit{str}01 \textit{I}4\textit{mm} structure is possible: the \V values of all the other \A configurations are too different to suit the experiment. On the other hand, in case of \C the broad distribution of \V  (Fig.~\ref{fig:LaC}) allows to include in addition to \textit{str}01 also structure \textit{str}03, and to some smaller extent also \textit{str}02, \textit{str}04, \textit{str}05, and \textit{str}06 can contribute to the spectrum. 

There are additional interactions, besides the electric quadrupole interaction, which could, in principle, contribute to the observed smearing-out of the features in \La spectrum of \C: the chemical shielding (including Knight contribution in metals), nuclear dipole-dipole interaction, and indirect nuclear spin-spin interaction. However, we show in the following paragraph that these interactions are too weak to be responsible for the observed effects.

The nucleus is shielded from the external static magnetic field by the surrounding electrons, as described by the chemical shielding $\sigma$ and Knight shielding $K$ in case of conduction electrons in metals. The total shielding has isotropic part $\sigma_0 + K_0$, which in case of \La here is responsible for a uniform shift of the spectrum by approximately +200~kHz from the Larmor frequency, and anisotropic part $\Delta\sigma + \Delta K$, which we can estimate by employing our DFT calculations. Calculated anisotropic shieldings (evaluated as difference between shieldings for field in directions [001] and [100]), $\Delta\sigma = 70$~ppm and $\Delta K = -600$~ppm, which cause broadening of $\sim 30$~kHz in addition to effect of quadrupole interaction. Values, $\Delta\sigma = 230$~ppm and $\Delta K = 260$~ppm, calculated for \A structure yield similar value $\sim 28$~kHz. The broadening effect  due to the anisotropy of chemical and Knight shielding is thus relatively small compared to the quadrupole interaction for our \La spectra, and moreover, it is present in spectra of both samples, therefore, it cannot be responsible for broadening of smearing-out of spectral features for \C only. Next, the effect of dipolar interaction with surrounding nuclear moments can be calculated by direct summation within a sphere of radius 100~\AA, yielding estimate of $< 1.8$ kHz for \C and $< 2.0$ kHz for \A. Indirect nuclear spin-spin interaction, mediated by conduction electrons (Rudermann-Kittel interaction), cannot usually be observed directly, however, its value rarely exceeds a few kHz.\cite{RK1,RK2}
The effects of these interactions on smearing-out the \La spectrum are thus negligible compared to the effect of electric quadrupole interaction and its distribution of \V. Therefore, the only possible interpretation of the smeared-out shape of \La spectra in \C is a presence of several different lanthanum sites with different \V. 

\begin{figure}
	\includegraphics[width=\columnwidth]{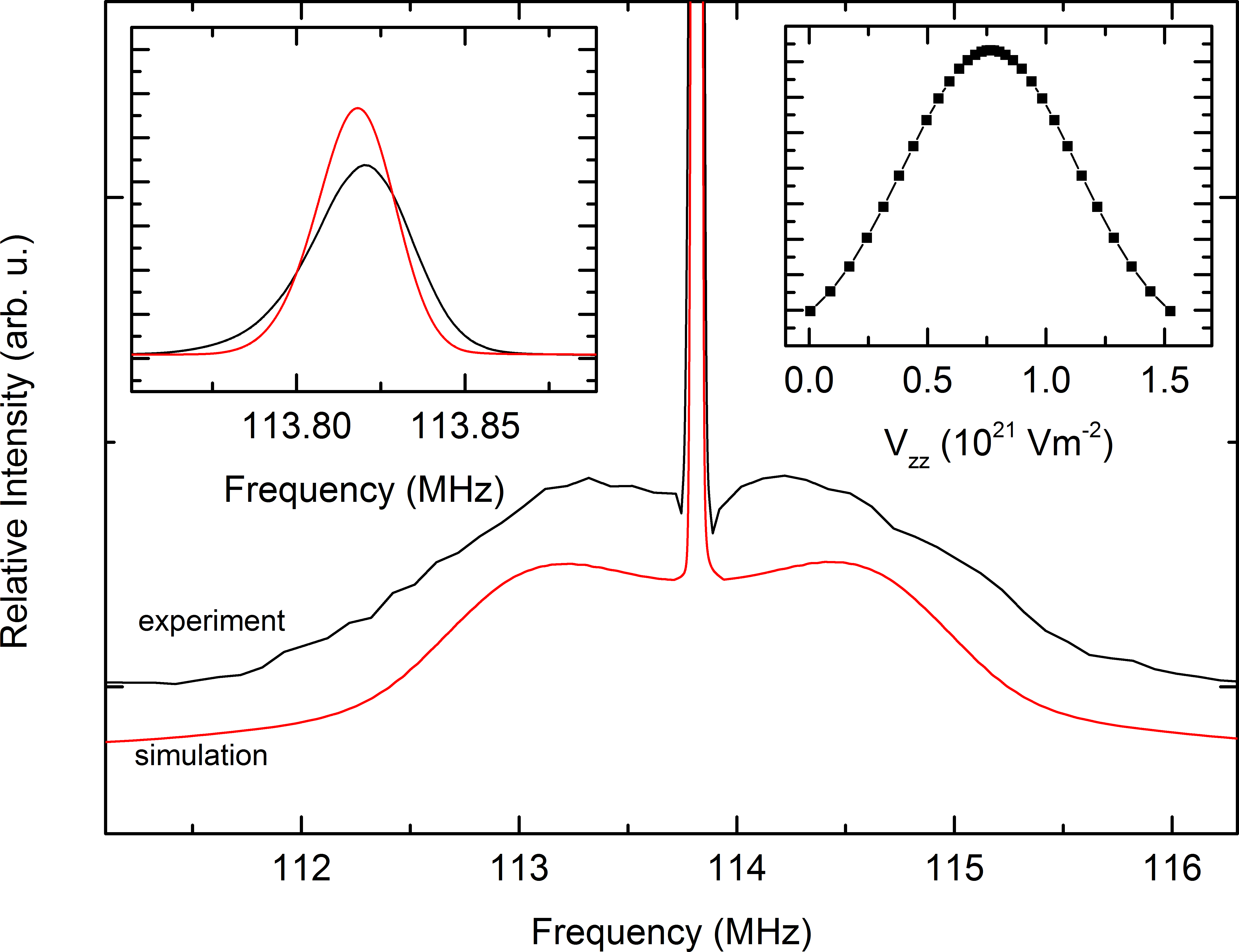}
	\caption{\label{fig:CuC} $^{65}$Cu NMR spectrum of \C in comparison with simulated spectrum. The inset on the left side details the central transition, inset on the right side shows distribution of \V parameter used in the simulation.}
\end{figure}

\begin{figure}
	\includegraphics[width=\columnwidth]{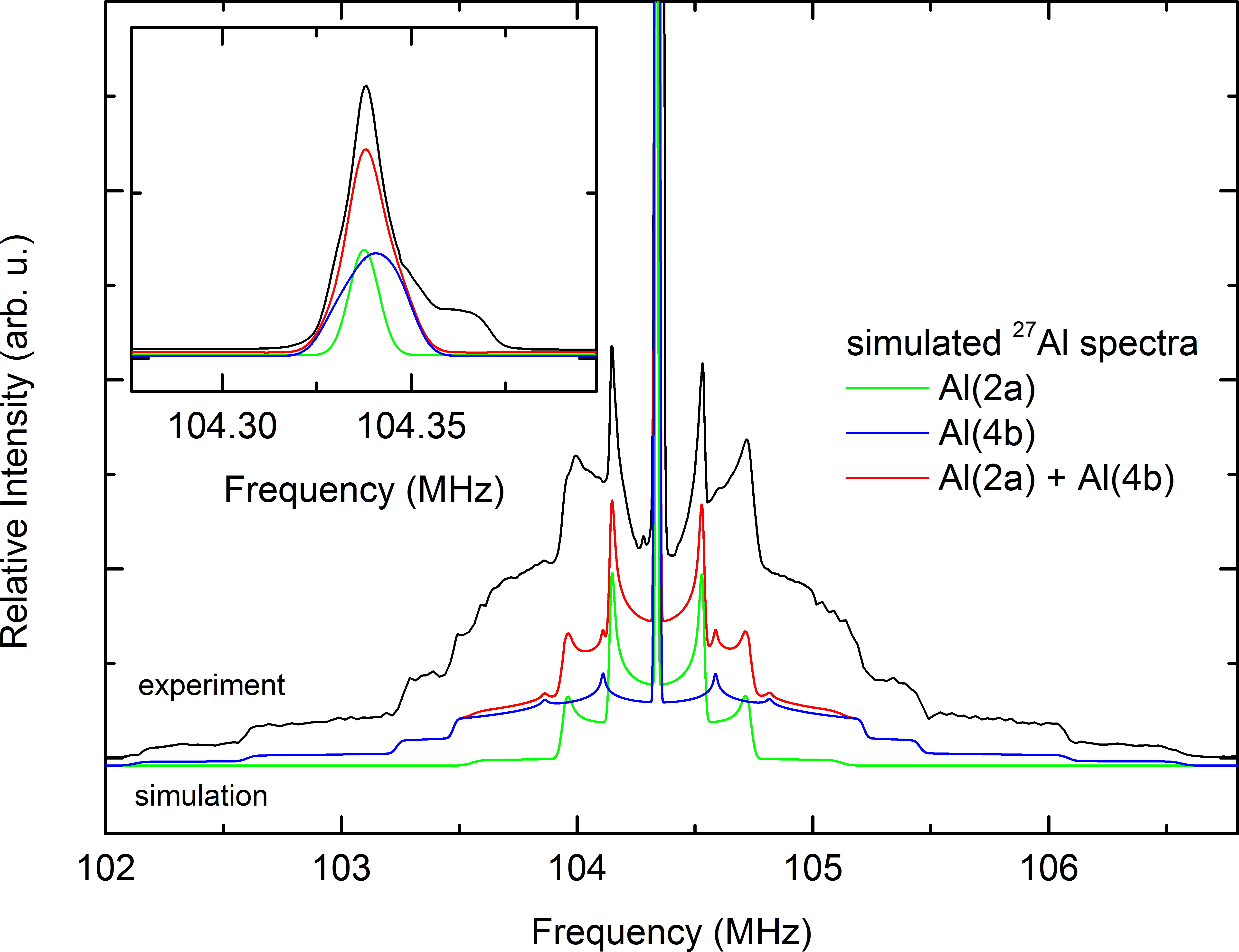}
	\caption{\label{fig:AlA} \Al NMR spectrum of \A in comparison with simulated spectrum. The inset shows the central transition.}
\end{figure}

\begin{figure}
	\includegraphics[width=\columnwidth]{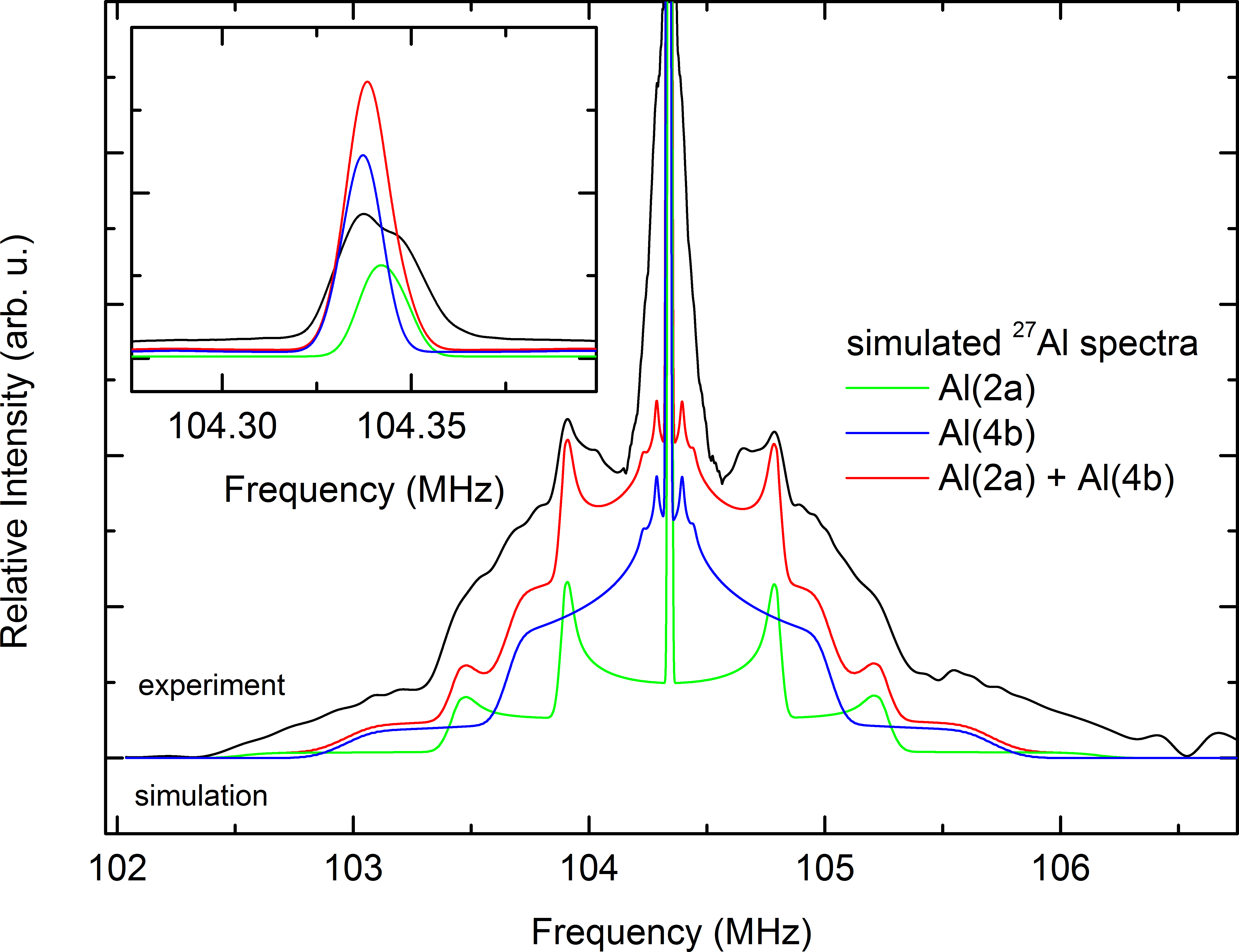}
	\caption{\label{fig:AlC} \Al NMR spectrum of \C in comparison with simulated spectrum. The inset shows the central transition.}
\end{figure}

Supplementary information concerning atomic ordering can be obtained from the other nuclei in the structure. $^{65}$Cu NMR spectrum in \C displays powder pattern, which is similarly unresolved as in case of \C \La spectrum (Fig. \ref{fig:CuC}): measured featureless powder pattern allows to match the spectrum with simulation using a very broad distribution of \V parameter. If the \C structure was well ordered, there would be  a single Cu(2a) site expected with a relatively sharp distribution of \V. Our calculations yield for such structure (\textit{str}01) $V_{zz} = 0.40 \times 10^{21}$ V/m$^{-2}$, which is a value apparently contributing to the experimental spectrum but clearly has to be accompanied by several other components with larger values of \V. Analogous spectrum for \A sample was not acquired due to a very low magnetogyric ratio of $^{197}$Au isotope. 

The notion about \C being disordered can be further supported by \Al NMR spectra of both samples (see Fig.~\ref{fig:AlA} and \ref{fig:AlC}). Al enters at least two nonequivalent sites even in the "ordered" $I4mm$ structure \textit{str01}, and above that, the Al(4b) sites lack local axial symmetry, i.e., $\eta$ can be non-zero and acts as another free parameter. Therefore, the approach of independently simulating the NMR spectra as applied for \La is more difficult to implement for \Al. For \A the most of the spectral features can still be captured by simulation using two \Al subspectra with ratio 1:2, which reflects the site occupancies Al(2a) and Al(4b), and relatively narrow distributions of \V around values $V_{zz}\mathrm{(2a)} = -0.74$ and $V_{zz}\mathrm{(4b)} = 2.09$ (in units $10^{21}$ V/m$^{-2}$), $\eta\mathrm{(2a)}=0$ due to symmetry and $\eta\mathrm{(4b)}= 0.57$. Despite the fact that the simulation did not consider distribution of $\eta\mathrm{(4b)}$, these values correspond relatively well to $V_{zz}\mathrm{(2a)} = -0.51$, $V_{zz}\mathrm{(4b)} = 1.91$, $\eta\mathrm{(2a)}=0$ and $\eta\mathrm{(4b)}= 0.83$ from \textit{str01} \A calculation. In case of \C the smeared-out features make an analogous simulation impractical and allow ambiguous results. Our attempt to capture the experimental spectrum of \C with two 1:2 Al subspectra (see Fig. \ref{fig:AlC}) apparently leaves space for additional spectral components and/or wide distributions of EFG parameters, thus pointing to presence of more than two nonequivalent Al sites in \C structure and again supporting our deduction that \C is at least partially disordered.

Experimental NMR spectra of all isotopes thus lead us to conclusion that a single atomic arrangement occurs in \A while at least two different configurations are present in \C compound. Situation in La compounds is  thus corroborating the scheme used to describe the specific heat data of Ce(Cu,Al)$_4$ series.\cite{Vlaskova}

\begin{figure}
	\includegraphics[width=\columnwidth]{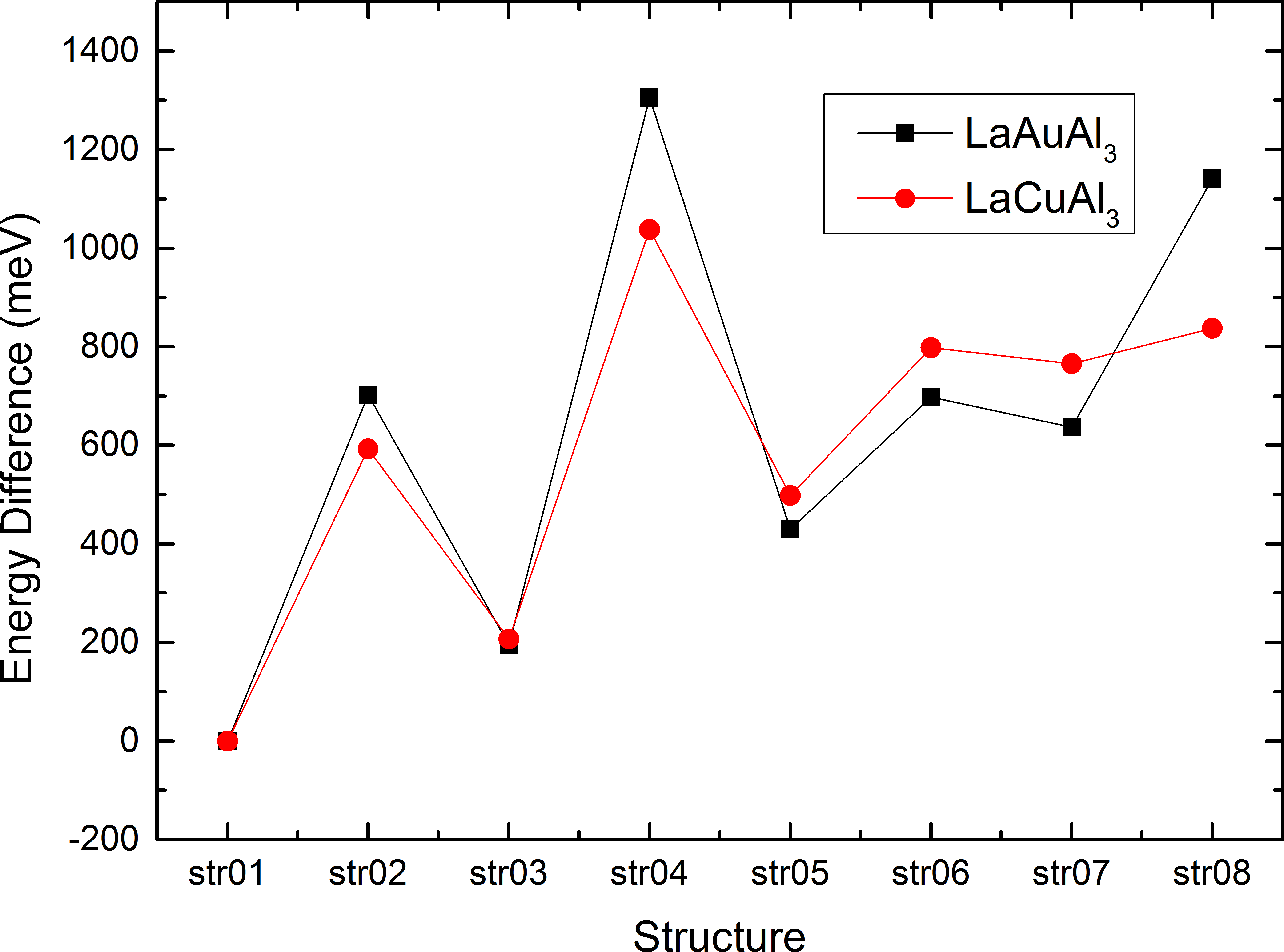}
	\caption{\label{fig:ene} Differences in calculated total energies for eight possible configurations in the stoichiometric structures of \A and \C.}
\end{figure}

\begin{figure}
	\includegraphics[width=\columnwidth]{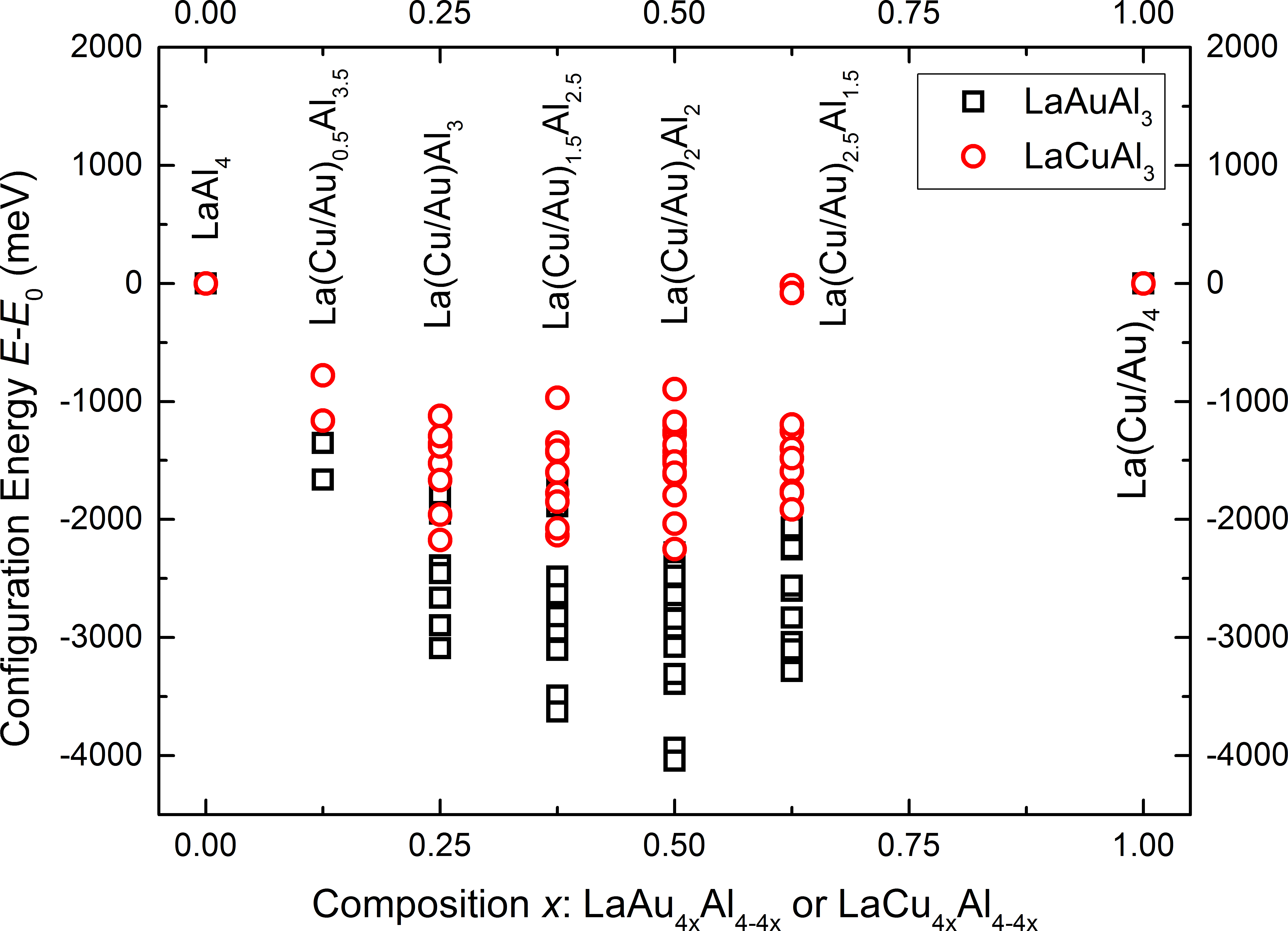}
	\caption{\label{fig:conf} Dependence of the difference between the total energy of structure with composition LaAu$_{4x}$Al$_{4-4x}$ (or LaCu$_{4x}$Al$_{4-4x}$) and the total energy of its constituents LaAl$_4$ and LaAu$_4$ (or LaCu$_4$, respectively).}
\end{figure}

Natural question then arises why the behavior of \C and \A is so different, i.e., why \C is more disordered than \A. In stoichiometric 1:1:3 cases for both \A and \C, $str$01 is the preferred arrangement of Au/Al and Cu/Al atoms, and the energies of the remaining structures $str02$--$str08$ are very similar (Fig.~\ref{fig:ene}). Apparently, to explore in more detail the differences in ordering, our model consisting of a single unit cell is too limiting. A supercell approach going beyond one cell would be very cumbersome, therefore, we instead calculated and fully optimized unit cells with varying stoichiometries: unit cells with several contents $x$ of Au (or Cu) between 0 and 1, i.e., from LaAl$_4$ to LaAu$_4$ (LaCu$_4$). We can now evaluate the stability of a given structure LaAu$_{4x}$Al$_{4-4x}$ by confronting its calculated total energy $E$ with the sum of calculated total energies of its constituents $E_0(x) = x E(\mathrm{LaAu}_4) + (1-x) E(\mathrm{LaAl}_4)$. Since there are again more possible arrangements of Au/Al for $x$ (especially for $x$ near 0.5) we observe a spread of energies for each value of $x$, see Fig.~\ref{fig:conf}. For $x< 0.5$ the most energetically favorable structure is always of the BaNiSn$_3$ type. Both \A and \C apparently are stable, however the minimum of the dependence in Fig.~\ref{fig:conf} is deeper for Au containing compounds. In general, for such mixtures the larger is the negative difference of $E-E_0$, the higher is the tendency of such mixture for ordering -- in order to maximize the number of interactions between different atoms.
 
This behavior of \A and \C can be explained by difference of atomic sizes and electronegativities between Au and Cu atoms. We evaluated both properties by "Atoms in Molecules" approach (Bader analysis \cite{bader}), where the atom is defined as a volume delimited by critical points (local minima or saddle points) in the charge density	. Calculated volume 24.4 \AA$^3$ of Cu atom in \C is rather close to 19.2 and 16.0 \AA$^3$ of aluminum in Al(2a) and Al(4b), respectively, and thus the mixing on individual atomic sites is relatively easy. On the other hand, gold atoms are somewhat larger (35.0 \AA$^3$) which prevents site mixing with smaller aluminum atoms (18.1 and 14.4 \AA$^3$) in the \A structure. Another point is higher difference between the electronegativity of Au and Al compared to such difference between Cu and Al, which can be best demonstrated using calculated valences (evaluated as the difference between atomic number $Z$ and the charge integrated within the atomic volume). In \A gold has valence $-2.86$ and aluminium sites Al(2a) and Al(4b) possess  $+0.01$ and $+0.75$, whereas in \C copper has valence $-1.91$ and both Al have $-0.22$ and $+0.44$. Larger difference between Au and Al valences thus again increases tendency for ordered arrangement.

It is worthwhile to mention certain analogy which can be seen when inspecting the binary phase diagrams. One can find several compounds with defined exact stoichiometry in the Au-Al binary phase diagram.\cite{diagrAu} On the other hand, Cu-Al phase diagram\cite{diagrCu} does not contain any line stoichiometric compound but rather regions of stability of several Cu-Al phases which allow for varying Cu-Al fractions. This observation reminds closely our case when comparing \C and \A.

\section{Conclusions}
The character of nuclear magnetic resonance spectra of \La, \Al, and $^{65}$Cu measured in \A and \C powder samples is dominated by electric quadrupole interaction. By simulating the NMR spectra and comparing to experiment, the spectral parameters were extracted and confronted with values obtained from the electronic structure calculations. In case of \A the \La spectrum can be interpreted by a single spectral component corresponding to uniform environment of La atoms in the structure, whereas in case of \C the spectrum decomposition yields a wide distribution of spectral parameters, and thus corresponds to multiple non-equivalent La positions in the structure. This scenario is also reflected in $^{65}$Cu NMR spectrum of \C and in \Al spectra of both samples, and together with analysis of our calculations leads to conclusion that \C is somewhat disordered compared to ordered \A.

\begin{acknowledgments}
This work was supported by the Czech Science Foundation under Grant No. 17-04925J, and by the project SVV-260442. Computational resources were provided by the CESNET LM2015042 and the CERIT Scientific Cloud LM2015085, provided under the programme "Projects of Large Research, Development, and Innovations Infrastructures".
\end{acknowledgments}

\section*{References}

\end{document}